\documentclass[10pt]{article}
\usepackage{hyperref,graphicx,amsthm,amsfonts,amsmath,authblk,amssymb,mathrsfs,amsthm,latexsym,amssymb,amsfonts,bm}
\usepackage{enumerate}
\usepackage{subfigure}
\usepackage[sort&compress,comma,numbers]{natbib}
\usepackage[top=2cm, bottom=2cm, outer=2cm, inner=2cm, heightrounded]{geometry}
\usepackage{color}
\usepackage{ulem}

\newcommand{\tr}{{\rm Tr}}
\newcommand{\sgn}{\mathrm{sgn}}

\renewcommand{\d}{\mathrm{d}}

\newcommand{\he}{H_{\rm eff}}

\newcommand{\piphi}{\pi_\phi}

\newcommand{\cc}{\mathscr{C}}
\newcommand{\hh}{\bm{H}}
\begin{document}
\title{Alternative dynamics in loop quantum Brans-Dicke cosmology}
\author[1]{Shupeng Song\thanks{songsp@mail.bnu.edu.cn} }
\author[1,2]{Cong Zhang\thanks{zhang.cong@mail.bnu.edu.cn} }
\author[1]{Yongge Ma\thanks{corresponding author: mayg@bnu.edu.cn}}

\affil[1]{ Department of Physics, Beijing Normal University, Beijing 100875, China}
\affil[2]{Faculty of Physics, University of Warsaw, Pasteura 5, 02-093 Warsaw, Poland}
%
\maketitle
\begin{abstract}
To inherit more features of full loop quantum Brans-Dicke theory, the Euclidean and Lorentzian terms of the Hamiltonian constraint are quantized independently in loop quantum Brans-Dicke cosmology. An alternative Hamiltonian constraint operator and its effective expression are obtained in the cosmological model. A residual quantum correction term is found in the effective Hamiltonian constraint, which has no analog in the effective Hamiltonian of the loop quantum cosmology from general relativity.
The dynamics driven by this effective Hamiltonian constraint is analyzed in detail. For the physically interesting case of $\omega\gg 1$, this effective Hamiltonian drives a bouncing evolution which evolves from a de Sitter universe to a classical Brans-Dicke solution.

\end{abstract}
\section{Introduction}
 How to unify general relativity (GR) with quantum mechanics by a theory of quantum gravity is a great challenge to theoretical physics. As a nonperturbative approach to quantum gravity, loop quantum gravity (LQG) has made remarkable progress in the past thirty years \cite{ashtekar2004back,rovelli2005quantum,thiemann2007modern,han2007fundamental} . According to LQG, spacetime consists of fundamental units of spacetime quanta since the spectra of the operators corresponding to the classical length, area and volume turned out to be discrete \cite{rovelli1994physical,ashtekar1997quantum,ashtekar1997quantumII,thiemann1998length,ma2010new,yang2016new}.
Despite these achievements, the dynamics of LQG is still an open issue, as the problem of how to suitably quantize and solve the Hamiltonian constraint is still unsolved.  There are some attempts to quantize the Hamiltonian constraints \cite{thiemann1998quantum,thiemann2006phoenix,han2006master,yang2015new,Domagala2015kse,alesci2015hamiltonian}, and some properties of the resulted operators are studied \cite{bonzom2011hamiltonian,alesci2013matrix,zhang2018towards,zhang2019bouncing}. The problems in the full LQG theory motivate us to consider the symmetry-reduced models, such as the homogeneous and isotropic cosmology,  on which the loop quantization method is applied \cite{bojowald2001absence,ashtekar2006quantumnature,ding2009effective}. The consequent quantum cosmology is called loop quantum cosmology (LQC).

 As a potential approach to address the dark energy and dark matter problems in the standard Lambda cold dark matter  model, a large variety of modified theories of gravity have been studied. Among these theories, a well-known one is the Brans-Dicke theory \cite{brans1961mach}, which is apparently compatible with Mach's principle. Loop quantization of this theory was studied in \cite{Zhang:2011vg}, where not only the kinematical Hilbert space but also the Hamiltonian constraint operator  were constructed. However, similar to the situation in LQG, it is still difficult to solve the Hamiltonian constraint in the full loop quantum Brans-Dicke theory (LQBDT). Then, the symmetry-reduced model of loop quantum Brans-Dicke cosmology (LQBDC) was developed afterward \cite{zhang2013loop,Artymowski:2013qua}. By solving the effective Hamiltonian constraint, one obtained a symmetric bouncing evolution of the Universe such that the classical big bang singularity was avoided in the quantum theory.

It should be noted that the Hamiltonian constraint in full LQG consists of two terms: the so-called Euclidean term and Lorentzian term. These two terms were first regularized and quantized as operators in \cite{thiemann1998quantum}. Classically the Lorentzian term is proportional to the Euclidean term in the spatially flat cosmological models. Thus one could combine the two terms into one term and then quantize it to obtain the Hamiltonian constraint operator in the cosmological models. In both standard LQC with massless scalar field and LQBDC, this treatment leads to the symmetric bounce  of the Universe \cite{bojowald2001absence,ashtekar2006quantumnature,zhang2013loop,Artymowski:2013qua}. Alternatively, the Lorentzian term could also be quantized independently  in the cosmological models by using Thiemann's trick as in full LQG and full LQBDT. This idea was first realized in \cite{yang2009alternative}, where an alternative Hamiltonian constraint operator was obtained in LQC. Notably, the effective Hamiltonian of this alternative operator was lately confirmed by the semiclassical analysis of Thiemann's Hamiltonian constraint operator in full LQG, which leads to an asymmetric bounce scenario in LQC \cite{assanioussi2018emergent,li2018towards}. This result relates the flat Friedmann-Lema\^{i}tre-Robertson-Walker cosmological spacetime with an asymptotic de Sitter spacetime. Thus an effective cosmological constant and an effective Newton constant were obtained in LQG \cite{li2018towards,assanioussi2018emergent}. This ambiguity also exists in LQBDC. To inherit more features of LQBDT, in this paper we will deal with the Euclidean and the Lorentzian terms independently in LQBDC. It will be shown that the main features of the effective dynamics of the alternative Hamiltonian in LQC are tenable by that of LQBDC.

The paper is arranged as follows. In Sec. \ref{se:two} the classical Brans-Dicke cosmology with  the coupling parameter $\omega\neq -3/2$ will be briefly reviewed, and then the kinematics of LQBDC will be introduced. In Sec. \ref{se:three}, the Hamiltonian constraint of the Brans-Dicke cosmological model will be quantized by using the strategy to treat the Euclidean and Lorentzian terms independently as in full LQBDT. In Sec. \ref{se:four}, the effective Hamiltonian constraint of the alternative Hamiltonian operator will be derived by the path-integral method in LQBDC. Then in Sec. \ref{se:five} the effective dynamics driven by the effective Hamiltonian will be studied. Finally, the results will be summarized and discussed in Sec. \ref{se:six}.

\section{Brans-Dicke cosmology and its loop quantization}\label{se:two}
The action of the original Brans-Dicke theory reads\cite{brans1961mach}
\begin{equation*}
  S[g,\phi]=\frac{1}{2\kappa} \int_{M} d^{4} x \sqrt{-g}\left[\phi R-\frac{\omega}{\phi}\left(\partial_{\mu} \phi\right) \partial^{\mu} \phi\right],
\end{equation*}
where $\kappa=8\pi G$ with $G$ the Newtonian gravitational constant, the scalar field $\phi$ is nonminimally coupled to the scalar curvature $R$, and the coupling constant $\omega$ is restricted by the observations to be bigger than $10^4$\cite{Will:2014kxa,will2018theory}.
In the connection formulation of Brans-Dicke theory, the phase space consists of canonical pairs of geometrical conjugate variables $(A_a^i, E^b_j )$ and scalar conjugate variables $(\phi,\Pi)$ , where $A_a^i$ is an SU(2) connection and $E^b_j$ is the densitized triad on the spatial manifold $M$. The nonvanished Poisson brackets between the canonical variables read
\begin{equation}
\begin{aligned}
\{A_a^i(x),E^b_j(y)\}&=\kappa\gamma\delta_a^b\delta^i_j\delta(x,y),\\
\{\phi(x),\Pi(y)\}&=\delta(x,y),
\end{aligned}
\end{equation}
where $\gamma$ is the Barbero-Immirzi parameter. In the case of the coupling constant $\omega\neq -3/2$ as required by the observation, the Hamiltonian constraint in Brans-Dicke theory reads \cite{zhang2013loop}
  \begin{equation}\label{eq:Hamiltonainfull}
  \begin{aligned}
  H=&\frac{\phi}{2}\left(F ^j_{ab}-(\gamma^2+\frac{1}{\phi^2})\epsilon_{jmn}\tilde{K}^m_a\tilde{K}^n_b\right)\frac{\epsilon_{jkl}E^a_kE^b_l}{\sqrt{q}}\\
  &+\frac{1}{3+2\omega}\left(\frac{(\tilde{K}^i_aE^a_i)^2}{\phi\sqrt{q}}+2\kappa\frac{(\tilde{K}_a^iE^a_i)\Pi}{\sqrt{q}}+\kappa^2\frac{\Pi^2\phi}{\sqrt{q}}\right) +\frac{\omega}{2\phi}\sqrt{q}(D^a\phi) D_a\phi+\sqrt{q}D_aD^a\phi=0,
  \end{aligned}
  \end{equation}
where $F^i_{ab}= 2\partial_{\left[a\right.}A^i_{\left.b\right]}+{\epsilon^i}_{kl}A^k_a A^l_b$ is the curvature of the connection $A_a^i$, $\tilde{K}^i_a$ is defined in \cite{zhang2013loop} , and $q$ is the determinant of physical 3-metric on $M$.

We will restrict ourselves to spatially flat, homogeneous and isotropic cosmology with the symmetry of $\mathcal{S}=\mathbb{R}^3\rtimes_\rho $SO(3). Then the spatial 3-manifold $M$ is diffeomorphic to $\mathbb{R}^3$. As in the standard treatment of LQC, we first introduce an ``elementary cubic cell" $\mathcal{V}$ on $M$ and restrict all integrals to this cell. Fix a fiducial 3-metric $\mathring{q}_{ab}$ and denote the volume of $\mathcal{V}$ measured with $\mathring{q}_{ab}$ by $V_0$.  Let $\mathring{e}^a_i$ and $\mathring{\omega}_a^i$ be the triad and cotriad adapted to $\cal{V}$ and satisfying $\mathring{\omega}_a^i\mathring{e}^b_i=\delta_a^b$ and $\mathring{q}_{ab}=\delta_{ij}\mathring{\omega}_a^i\mathring{\omega}_b^j$. By fixing the local diffeomorphism and internal gauge freedom, the basic variables are reduced to
\begin{equation}
A_a^i=c V_0^{-1/3}\mathring{\omega}_a^i,~E^b_j=pV_0^{-2/3}\sqrt{\mathring{q}}\mathring{e}^b_j,~\Pi=V_0^{-1}\sqrt{\mathring{q} }\piphi.
\end{equation}
The nontrivial Poisson brackets among reduced variables $c$, $b$, $\phi$, and $\piphi$ read
\begin{equation}
\{c,p\}=\frac{\kappa\gamma}{3},~\{\phi,\piphi\}=1.
\end{equation}
The remaining Hamiltonian constraint \eqref{eq:Hamiltonainfull} is reduced to
\begin{equation}\label{eq:Hamiltonianc}
  H=-\frac{3c^2\sqrt{|p|}}{\gamma^2\phi}+\frac{1}{(3+2\omega)\phi |p|^{3/2}}\left(\frac{3cp}{\gamma}+\kappa\piphi\phi\right)^2=0.
  \end{equation}
The kinematical Hilbert space $\mathcal H$ of the LQBDC can be given by the direct product of the geometric sector $\mathcal{H}_{\rm geo}=L^2(\mathbb{R}_{\rm Bohr},\d\mu_H)$ \cite{ashtekar2003mathematical,thiemann2007modern}, where $\mathbb{R}_{\rm Bohr}$ is the Bohr compactification of $\mathbb{R}$ and ${\mathrm d}\mu_{\mathrm{Bohr}}$ is the Haar measure, and the scalar field sector $\mathcal{H}_{\rm sca}=L^2(\mathbb{R},\d\mu)$, which is the usual Schr\"{o}dinger representation, i.e.,
 \begin{equation}
 \mathcal{H}=L^2(\mathbb{R}_{\rm Bohr},\d\mu_H)\otimes L^2(\mathbb{R},\d\mu).
 \end{equation}
In $\mathcal{H}_{\rm sca}$, one has the configuration operator $\hat{\phi}$ defined as multiplication and the momentum operator $\hat{\pi}_{\phi}:=i\hbar\d / \d \phi$. The generalized eigenstates $|\phi)$ of $\hat{\phi}$ contribute a generalized basis of $\mathcal{H}_{\rm sca}$.
In $\mathcal{H}_{\rm geo}$, there are two fundamental operators, namely the momentum operator $\hat{p}$ which represents the area of each side of $\mathcal{V}$ and the configuration operator ${\widehat{\exp{({\rm i}\lambda c)}}}$ which represents the holonomy of the reduced connection $c$ along an edge parallel to an edge of $\mathcal{V}$.
Since we will follow the improved scheme as in \cite{ashtekar2006quantumnature}, it is convenient to introduce a new operator
$$
\hat{v}=\frac{\textrm{sgn}(\hat{p})|\hat{p}|^{3/2}}{2\pi\gamma\ell^2_{p}\sqrt\Delta},
$$
where $\ell_{p}=\sqrt{G\hbar}$ is the Planck length and $\Delta=4\sqrt{3}\pi\gamma \ell_p^2$ denotes the area gap in full LQBDT. Note that $\hat{v}$ is actually a dimensionless variable representing the physical volume of $\mathcal{V}$. The eigenstates $|v\rangle$ of the operator $\hat{v}$ are labeled by real numbers $v$ and contribute an orthonormal basis in $\mathcal{H}_{\rm geo}$ such that
\begin{align}
\langle v|v'\rangle=\delta_{v,v'}\, ,
\end{align}
where $\delta_{v,v'}$ is the Kronecker delta. A general state in ${\mathcal H}_{\rm geo}$  can be expressed as a countable sum: $|\psi\rangle=\sum\psi_n|v_n\rangle$ and thus the inner product reads
$$\langle\psi^{(1)}|\psi^{(2)}\rangle=\sum_n\overline{\psi^{(1)}_n}\psi^{(2)}_n.$$
  It should be noted that the operator which measures the physical volume $V$ of $\mathcal{V}$ is given by
\begin{equation}
\hat{V}=2\pi\gamma\ell_{p}^2\sqrt{\Delta}\,|\hat v|.
\end{equation}
where $|\hat{v}|$ is the absolute value of the operator $\hat{v}$.
One prefers to use the holonomy operator $\widehat{e^{{\rm i}b/2}}$, where $b:=\bar{\mu}c$ with $\bar\mu=\sqrt{\Delta/|p|}$.
Note that $\widehat{e^{{\rm i}b/2}}$ represents the holonomy $h^{(\bar\mu)}_i$ of $c$ along an edge parallel to the  triad $\mathring{e}^a_i$ whose length with respect to the physical metric is $\sqrt{\Delta} $. Thus the edge underlying $h^{(\bar\mu)}_i$ takes the minimal length of the quantum geometry. The variables $b$ and $v$ are conjugate to each other, since
$$\{b,v\}=\frac{2}{\hbar}.$$ Hence one has
\begin{equation}
\widehat{e^{{\rm i}b/2}}\,|v\rangle=|v+1\rangle.
\end{equation}
Actually, the holonomy operator $\widehat{h}_i^{(\bar\mu)}$ can be expressed as
\begin{equation}\label{eqn:i-holonomy}
\begin{aligned}
&\widehat{h}^{(\bar\mu)}_i=\frac{1}{2}\left(\widehat{e^{{\rm i}b/2}}+\widehat{e^{-{\rm i}b/2}}\right)-{\rm i}\left( \widehat{e^{{\rm i}b/2}}-\widehat{e^{-{\rm i}b/2}}\right)\tau_i,
\end{aligned}
\end{equation}
where $\tau_i$ are the generators of Lie algebra $\mathfrak{su}(2)$ \cite{ashtekar2006quantumnature}.

\section{Alternative Hamiltonian constraint operator}\label{se:three}
In the homogeneous cosmological model, the Hamiltonian constraint \eqref{eq:Hamiltonainfull} can be written as
\begin{equation}
\begin{aligned}\label{eq:Hamilton-c}
H=\frac{\phi}{2}\left(F^j_{ab}-(\gamma^2+\frac{1}{\phi^2})\epsilon_{jmn}\tilde{K}^m_a\tilde{K}^n_b\right)\frac{\epsilon_{jkl}E^a_kE^b_l}{\sqrt{q}} +\frac{1}{3+2\omega}\left(\frac{(\tilde{K}^i_aE^a_i)^2}{\phi\sqrt{q}} +2\kappa\frac{(\tilde{K}_a^iE^a_i)\Pi}{\sqrt{q}}+\kappa^2\frac{\Pi^2\phi}{\sqrt{q}}\right)=0.
 \end{aligned}
 \end{equation}
 Similar to the case of full LQBDT, there is no operator corresponding to the connection $A_a^i(x)$ in LQBDC. Hence, one has to express the curvature $F^j_{ab}$ in \eqref{eq:Hamilton-c} by holonomies. This can be accomplished by using Thiemann's tricks \cite{thiemann2007modern}. Classically the curvature in our cosmological model can be regularized on the elementary cell as \cite{ashtekar2006quantumnature}
 \begin{equation}
 F_{ab}^k=\lim_{\lambda\to 0} \tr\left(-2\frac{h_{ij}^{(\lambda)}\tau^k}{\lambda^2V_0^{2/3}}\right)\mathring{\omega}_a^i\mathring{\omega}_b^j,
 \end{equation}
where $h_{ij}^{(\lambda)}=h_i^{(\lambda)}h_j^{(\lambda)}(h_i^{(\lambda)})^{-1}(h_j^{(\lambda)})^{-1}$ is the holonomy around the loop formed by the two edges of $\cal{V}$ that are tangent to $e_i^a$ and $e_j^b$ whose length is $\lambda V^{1/3}$ with respect to the fiducial metric $\mathring{q}_{ab}$ respectively. To quantize the Hamiltonian constraint, we also need to use the regularizations
 \begin{equation}
\frac{\varepsilon^{ijk}E^b_jE^c_k}{\sqrt{\det(q)}}=\lim_{\lambda\to 0}\frac{2\sgn(p)\tr(h_m^{(\lambda)} \{(h_m^{(\lambda)})^{-1}),V\}\tau^i)}{\kappa\gamma\lambda V_0^{1/3}}\mathring{\omega}^m_a\varepsilon^{abc},
\end{equation}
and
\begin{equation}
\tilde{K}_a^i(x)=\frac{1}{2\gamma(\kappa\gamma)^2}\{A_a^i(x), \{C,V\}\},
\end{equation}
 where $\sgn(p)$ denotes the sign of $p$ and $C=\int \d^3x{\epsilon_i}^{jk} F_{ab}^i(x)E_j^a(x)E_k^b(x)/\sqrt{q(x)}$. The integration of the Hamiltonian \eqref{eq:Hamilton-c} reads
 \begin{equation}
 \cc=\int_{\mathcal{V}}\d^3 xH(x)=\lim_{\lambda\to 0}H^{(\lambda)}
 \end{equation}
where
\begin{equation}
\begin{aligned}
H^{(\lambda)}&= -\phi\frac{\sgn(p)}{2\pi G\gamma\lambda^3}\tr(h_{kj}^{(\lambda)}\tau^i)\tr(h_m^{(\lambda)}\{(h_m^{(\lambda)})^{-1},V\}\tau_i)\varepsilon^{kjm}\\
&+\frac{\sgn(p)}{\gamma^2(8\pi G\gamma)^5\lambda^3}\phi(\gamma^2+\frac{1}{\phi^2})\varepsilon^{ijk}\tr(h_i^{(\lambda)-1}\{h_i^{(\lambda)},\{C,V\}\}h_j^{(\lambda)-1}\{h_j^{(\lambda)},\{C,V\}\}h_k^{(\lambda)-1}\{h_k^{(\lambda)},V\})\\
&+\frac{1}{2\omega+3}\left(\frac{(\{C,V\})^2}{4\gamma^2(\kappa\gamma)^2\phi V }+\frac{\{C,V\}\piphi}{\gamma^2 V}+\kappa^2\frac{\piphi^2\phi}{V}\right).
\end{aligned}
\end{equation}
 However, the family of operators $\hat{H}^{(\lambda)}$ does not converge as $\lambda\to 0$. Thus, in the so-called $\bar\mu$-scheme \cite{ashtekar2006quantumnature}, one fixed the length $\lambda$ of the edge underlying the holonomies in the Hamiltonian to $\bar\mu=\sqrt{\Delta/|p|}$, which implies that the curvature is smeared over the elementary faces with the physical area Ar$_{\square}=\Delta$. By this treatment, we obtain the Hamiltonian constraint operator as
 \begin{equation}\label{eq:cc}
 \hat\cc=\lim_{\lambda\to \bar\mu}\hat{H}^{(\lambda)}.
 \end{equation}
 It should be noted that classically one has
 \begin{equation}
 \lim_{\lambda\to\bar\mu}\{h^{(\lambda)},\tilde{K}\}=\frac{2}{3}\{h^{\bar\mu},\tilde{K}\},
 \end{equation}
 where $\tilde{K}=\int\d^3 x\tilde{K}^i_aE^a_i$. Hence, in the expression of $\lim_{\lambda\to\bar\mu}$, the commutator $[\widehat{h}^{(\lambda)}, \hat{\tilde{K}}]$ would be replaced by $\frac{2}{3}[\widehat{h}^{\bar{\mu}}, \hat{\tilde{K}}]$.
It is convenient to split the expression of \eqref{eq:cc} into three parts as $\hat\cc=\hat{\cc}_1+\hat{\cc}_2+\hat{\cc}_3$. Their actions on the basis $|v, \phi\rangle=|v\rangle\otimes |\phi)$ of $\mathcal{H}$ are given by:
  \begin{equation}
 \begin{aligned}\label{eq:cc1}
\hat{\cc}_1|v,\phi\rangle=&\phi\sin(b)\hat A\sin(b)|v,\phi\rangle\\
=&\frac{1}{8}\alpha\phi \left( f_+(v)|v+4,\phi\rangle+f_0(v)|v,\phi\rangle+f_-(v)|v-4,\phi\rangle\right)
\end{aligned}
\end{equation}
with $\hat A=-i \hat v\left(\sin\frac{b}{2}|\hat v|\cos\frac{b}{2}-\cos\frac{b}{2}|\hat v|\sin\frac{b}{2}\right)$, $\alpha=6\pi\gamma\ell_p^2/\sqrt{\Delta}$,
$f_+(v)=(v +2) (\left| v +1\right| -\left| v +3\right| )$, $f_-(v)=f_+(v-4)$, and $f_0(v)=-f_+(v)-f_-(v)$,
\begin{equation}
\begin{aligned}\label{eq:cc2}
\hat{\cc}_2|v,\phi\rangle=&\frac{\alpha}{256\gamma^2}\phi(\gamma^2+\frac{1}{\phi^2})\hat\beta\hat A\hat\beta|v,\phi\rangle\\
=&-\frac{\alpha}{16^3\times 2\gamma^2}\phi(\gamma^2+\frac{1}{\phi^2})\Big(g^\Delta_+(v)A(v+4)g^\Delta_+(v+4)|v+8,\phi\rangle\\
 &-\big(g^\Delta_+(v)A(v+4)g^\Delta_-(v+4)+g^\Delta_+(v-4)A(v-4)g^\Delta_-(v)\big)|v,\phi\rangle\\
 &+g^\Delta_-(v-4)A(v-4)g^\Delta_-(v)|v-8,\phi\rangle\Big)
\end{aligned}
\end{equation}
 with $\hat\beta=2\left(\sin\frac{b}{2}[\hat c,|\hat v|]\cos\frac{b}{2}-\cos\frac{b}{2}[\hat c,|\hat v|]\sin\frac{b}{2}\right)$, $\hat c=2\sin(b)\hat A\sin(b)$,
 $g_+(v):=f_+(v)(|v|-|v+4|)$, $g_-(v):=f_-(v)(|v-4|-|v|)$, and $g_\pm^\Delta(v):=g_\pm(v+1)-g_\pm(v-1)$, and
\begin{equation}
\begin{aligned}\label{eq:cc3}
\hat{\cc}_3|v,\phi\rangle=&\frac{\alpha}{3+2\omega}\sqrt{|\widehat{v^{-1}}|}\left(\frac{-3[\hat c,|\hat{v}|]^2}{64\gamma^2 \phi }+\kappa \frac{3[\hat c,|\hat{v}|]\hat{\pi}_{\phi}}{4i\alpha\gamma\sqrt{\Delta}}+\kappa^2\frac{3}{2\alpha^2\Delta}(\hat{\pi}_{\phi}^2\hat{\phi}+\hat{\phi}\hat{\pi}_{\phi}^2)\right)\sqrt{|\widehat{v^{-1}}|}|v,\phi\rangle\\
=&(\frac{\sqrt{3}}{32\gamma})^2\frac{\alpha}{\phi}\Big( \frac{g_+(v)g_+(v+4)}{\sqrt{|v(v+8)|}}|v+8,\phi\rangle-\frac{g_+(v)g_-(v+4)+g_-(v)g_+(v-4)}{|v|}|v,\phi\rangle\\
&+\frac{g_-(v)g_-(v-4)}{\sqrt{|v||v-8|}}|v-8,\phi\rangle\Big)+\kappa\frac{3}{i 16\gamma\sqrt{\Delta}}  \piphi \Big(\frac{g_+(v)}{\sqrt{|v(v+4)|}}|v+4,\phi\rangle\\
&-\frac{g_-(v)}{\sqrt{|v(v-4)|}}|v-4,\phi\rangle\Big)+\kappa^2\frac{3}{2\alpha\Delta}\frac{1}{|v|}(\hat{\pi}^2_\phi\phi+\phi\hat{\pi}^2_\phi)|v,\phi\rangle,
\end{aligned}
\end{equation}
where $\widehat{v^{-1}}$ is defined by $\widehat{v^{-1}}|v\rangle=v^{-1}|v\rangle$ if $v\neq0$, and $\widehat{v^{-1}}|v\rangle=0$ if $v=0$\cite{assanioussi2017time}.

\section{The effective Hamiltonian constraint}\label{se:four}
To get an effective Hamiltonian constraint, we calculate the transition amplitude of the Hamiltonian constraint operator \eqref{eq:cc} as
\begin{equation}\label{eq:tran-ampl}
A(v_f,\phi_f;v_i,\phi_i)=\langle v_f,\phi_f|v_i,\phi_i\rangle_{\rm phy}=\lim_{\alpha_0\to \infty}\int_{-\alpha_0}^{\alpha_0}\d\alpha\langle v_f\phi_f|e^{i\alpha\hat{\cc}}|v_i,\phi_i\rangle.
\end{equation}
Dividing the path into $N$ parts by setting $\alpha=\sum_{n=1}^N \epsilon_n$ and inserting the basis, we have
\begin{equation}
\langle v_f,\phi_f|e^{i\alpha \hat{\cc}}|v_i,\phi_i\rangle=\sum_{v_{N-1},\cdots,v_1}\int\d \phi_{N-1}\cdots\d\phi_1\prod_{n=1}^N\langle \phi_n,v_n|e^{i\epsilon_n\hat{\cc}}|v_{n_1},\phi_{n-1}\rangle,
\end{equation}
where $\langle \phi_n,v_n|e^{i\epsilon_n\hat\cc}|v_{n-1},\phi_{n-1}\rangle$ can be calculated by using the formula
\begin{equation}
\int\d\phi_n\langle \phi_n,v_n|e^{i\epsilon_n\hat{\cc}}|\phi_{n-1},v_{n-1}\rangle=\delta_{v_n,v_{n-1}}-i\epsilon_n\int\d\phi_n\langle\phi_n,v_n|(\hat{\cc}_1+\hat\cc_2+\hat\cc_3)|v_{n-1},\phi_{n-1}\rangle.
\end{equation}
By Eqs. \eqref{eq:cc1}--\eqref{eq:cc3}, we obtain
\begin{equation}
\begin{aligned}
&\int d\phi_n\langle\phi_n,v_n|\cc_1|v_{n-1}\phi_{n-1}\rangle&\\
=&-\frac{1}{2\pi\hbar}\frac{\alpha}{8}\int \d\phi_n \int d\pi_n e^{i\frac{\pi_n}{\hbar}(\phi_n-\phi_{n-1})}\phi_n  (v_n+v_{n+1})(\delta_{v_n,v_{n-1}+4}-2\delta_{v_n,v_{n-1}}+\delta_{v_n,v_{n-1}-4}),\\
&\int d\phi_n\langle\phi_{n},v_{n}|\hat\cc_2|v_{n-1}\phi_{n-1}\rangle&\\
 =&\frac{1}{2\pi\hbar}\frac{\alpha}{32\gamma^2}\int \d\phi_n \int d\pi_n e^{i\frac{\pi_n}{\hbar}(\phi_n-\phi_{n-1})}\phi_n(\gamma^2+\frac{1}{\phi_n^2}) (v_n+v_{n-1})( \delta_{v_n,v_{n-1}+8}-2\delta_{v_{n-1},v_n}+\delta_{v_n,v_{n-1}-8}),
\end{aligned}
\end{equation}
and
\begin{equation}
\begin{aligned}
&\int \d\phi_n\langle \phi_n v_n|\hat\cc_3|\phi_{n-1},v_{n-1}\rangle\\
=&\frac{1}{3+2\omega}\left(-\frac{1}{2\pi\hbar}(\frac{\sqrt{3 }}{4\gamma})^2\int \d\phi_n\int\d\pi_n e^{i\frac{\pi_n}{\hbar}(\phi_n-\phi_{n-1})}\frac{\alpha}{\phi_n}(\sqrt{v_{n}v_{n-1}}+\frac{4}{\sqrt{v_n v_{n+1}}})\Big( \delta_{v_n,v_{n-1}+8}-2\delta_{v_{n-1},v_n}+\delta_{v_{n-1}-8,v_n}\Big)\right.\\
-&\frac{1}{2\pi\hbar}(\frac{\sqrt{3}}{4\gamma})^2\int \d\phi_n\int\d\pi_n e^{i\frac{\pi_n}{\hbar}(\phi_n-\phi_{n-1})}\frac{\alpha}{\phi_n}\frac{8}{\sqrt{v_{n}v_{n-1}}}\Big( \delta_{v_n,v_{n-1}+8}+\delta_{v_{n-1}-8,v_n}\Big)\\
+&\frac{1}{2\pi\hbar}\kappa\frac{3}{i\gamma 4\sqrt{\Delta}}\int\d\phi_n\int\d\pi_n e^{i\frac{\pi_n}{\hbar}(\phi_n-\phi_{n-1})}\pi_n \frac{v_n+v_{n+1}}{\sqrt{v_n(v_{n+1})}} (\delta_{v_n,v_{n-1}+4}-\delta_{v_n,v_{n-1}-4})\\
&+\left.\frac{1}{2\pi\hbar}\kappa^2\frac{3}{2\alpha\Delta}\int\d\phi_n\int\d\pi_n(\phi_n+\phi_{n+1})\pi_n^2 e^{\frac{\pi_n}{\hbar}(\phi_n-\phi_{n-1})} \frac{1}{v_{n-1}}\delta_{v_n,v_{n-1}}\right).
\end{aligned}
\end{equation}
Combining these equations and the formulas
\begin{equation*}
\begin{aligned}
\delta_{v_n,v_{n-1}+4}-2\delta_{v_n,v_{n-1}}+\delta_{v_n,v_{n-1}-4}&=-\frac{1}{\pi}\int_0^{\pi }\d b_n 4 e^{-i b_n (v_n-v_{n-1}) } \sin ^2(b_n), \\
\delta_{v_n,v_{n-1}+4}-\delta_{v_n,v_{n-1}-4}&=\frac{i}{\pi}\int_0^{\pi} \d b_n 2 e^{-i b_n(v_n-v_{n-1}) } \sin (2 b_n),\\
\delta_{v_n,v_{n-1}}&=\frac{1}{\pi}\int_0^{\pi }\d b_n e^{-i \frac{1}{2}b_n (v_n-v_{n-1}) },
\end{aligned}
\end{equation*}
we get
\begin{equation}
\begin{aligned}
&\langle \phi_n v_n|\hat\cc|\phi_{n-1},v_{n-1}\rangle\\
=&\frac{1}{2\pi\hbar}\int\d\pi_n e^{i\frac{\pi_n}{\hbar}(\phi_n-\phi_{n-1})} \frac{1}{\pi}\int_0^{\pi} \d b_n  e^{-i\frac{1}{2} b_n (v_n-v_{n-1}) }\left(
\frac{\alpha}{8}\phi_n  (v_n+v_{n+1})4\sin ^2(b_n) -\frac{\alpha}{32\gamma^2}\phi_n(\gamma^2+\frac{1}{\phi_n^2}) (v_n+v_{n-1}) 4\sin ^2(2b_n)\right. \\
&+\frac{1}{3+2\omega}\Big((\frac{\sqrt{3}}{4\gamma})^2\frac{\alpha}{\phi_n}(\sqrt{v_{n}v_{n-1}}+\frac{4}{\sqrt{v_n v_{n+1}}})4 \sin ^2(2b_n) -(\frac{\sqrt{3}}{4 \gamma })^2\frac{\alpha}{\phi_n}\frac{8}{\sqrt{v_{n}v_{n-1}}}2 \cos (4 b_n) +\\
&+\left.\kappa\frac{3}{4\gamma\sqrt{\Delta}\alpha}\pi_n \frac{v_n+v_{n+1}}{\sqrt{v_n(v_{n+1})}} 2 \sin (2 b_n)+\kappa^2\frac{3}{2\alpha\Delta}(\phi_n+\phi_{n+1})\pi_n^2 \frac{1}{v_{n-1}}\Big)\right).
\end{aligned}
\end{equation}
Hence the transition amplitude \eqref{eq:tran-ampl} can be expressed as
\begin{equation}\label{eq:tran-ampl2}
\begin{aligned}
&A(v_f,\phi_f;v_i,\phi_i)\\
=&\lim_{\alpha_0\to \infty}\int_{-\alpha_0}^{\alpha_0}\d \alpha_0\lim_{N\to \infty} \sum_{\{v_{N-1},\cdots,v_1\}}\int d\phi_{N-1}\cdots \d\phi_1\prod_{n=1}^{N}\langle \phi_n,v_n|e^{-i\epsilon_n C}|\phi_{n-1},v_{n-1}\rangle\\
=&\int\mathcal{D} \alpha \int \mathcal{D}\phi \int\mathcal{D}\pi \int\mathcal{D} b\int\mathcal{D} v \exp\Big\{\frac{i}{\hbar}\int \d\tau\left(  \pi\dot{\phi}-\frac{\hbar}{2}  b\dot{v}\right. +\hbar\Bigg(\alpha\phi  v \sin ^2(b) -\frac{\alpha}{4\gamma^2}\phi(\gamma^2+\frac{1}{\phi^2}) v \sin ^2(2b) \\
&-\frac{1}{3+2\omega}\left.\Big((\frac{\sqrt{3} }{ \gamma})^2\frac{1}{\phi}(v+\frac{4}{v}) \sin ^2(2b)- (\frac{\sqrt3 }{ \gamma})^2\frac{1}{\phi v} \cos (4  b) +\kappa\frac{ 3}{\gamma\sqrt{\Delta}}\pi_\phi \sin (2 b)+\kappa^2\frac{3 }{\alpha\Delta}\phi\piphi^2 \frac{1}{v}\Big)\Bigg)\right)\Big\}.
\end{aligned}
\end{equation}
Therefore, the effective Hamiltonian constraint can be read from Eq.~\eqref{eq:tran-ampl2} as
\begin{equation}\label{eq:effh}
\begin{aligned}
\he=&-\alpha\phi  v \sin ^2(b) +\frac{\alpha}{4\gamma^2}\phi(\gamma^2+\frac{1}{\phi^2}) v \sin ^2(2b) +\frac{1}{3+2\omega}\frac{\alpha}{\phi v}\Big(\frac{\sqrt 3}{2\gamma} v\sin(2b)+\frac{\sqrt3 \kappa}{\alpha\sqrt\Delta}\phi\pi_\phi\Big)^2\\
&-\frac{3\alpha}{3+2\omega}\frac{1}{\gamma^2v\phi}(\cos(4b)-\sin^2(2b)).
\end{aligned}
\end{equation}
In the limit $b\to 0$, we have
\begin{equation}\label{eq:effapp}
\begin{aligned}
H_{\rm eff}=&-\frac{\alpha}{\phi\gamma^2}v b^2+\frac{1}{3+2\omega}\frac{\alpha}{\phi v}\Big(\frac{\sqrt 3}{\gamma} vb+\frac{\sqrt3 \kappa}{\alpha\sqrt\Delta}\phi\pi_\phi\Big)^2-\frac{3\alpha}{3+2\omega}\frac{1}{\gamma^2v\phi}(1-12b^2).
\end{aligned}
\end{equation}
Equation \eqref{eq:effapp} is different from the classical Brans-Dicke Hamiltonian constraint \eqref{eq:Hamiltonianc} by the residual term $\frac{3\alpha}{3+2\omega}\frac{1}{\gamma^2v\phi}(1-12 b^2)$. In order to compare this term with the others, it is convenient to introduce a new variable
\begin{equation}
B=\frac{b}{4\pi G\gamma\sqrt\Delta},
\end{equation}
 which is canonically conjugate to the physical volume $V$ of the elementary cell $\mathcal{V}$ as
\begin{equation}
\{B,V\}=1.
\end{equation}
Then Eq. \eqref{eq:effapp} can be reexpressed in terms of $B$ and $V$ as
\begin{equation}\label{eq:H-BV}
\he= -\frac{3\kappa^2}{4\phi}VB^2+\frac{\kappa^2}{3+2\omega}\frac{1}{\phi V}(\frac{3}{2}BV+\piphi\phi)^2-\frac{\hbar^2}{3+2\omega}\frac{9\kappa^2}{16V\phi}(1-3\kappa^2\gamma^2\Delta B^2).
\end{equation}
It is obvious from Eq.\eqref{eq:H-BV} that the residual term in \eqref{eq:effapp} is of order $\hbar^2$, which is certainly a quantum correction. By checking the derivation procedure of the effective Hamiltonian, one can find that the residual term comes from the effect of $[\hat c,|\hat{v}|]^2$ in $\hat{\cc}_3$. Thus this is a particular term existing in the effective theory of LQBDC, since there is no square term of a commutator in the expression of the Hamiltonian constraint operator in the usual LQC. For semiclassical consideration, one may get rid of this term and obtain the following effective Hamiltonian constraint
\begin{equation}\label{Hamiltonian-e}
\he=-\alpha\phi  v \sin ^4(b) +\frac{\alpha}{4\gamma^2}\phi(\gamma^2+\frac{1}{\phi^2}) v \sin ^2(2b) +\frac{1}{3+2\omega}\frac{\alpha}{\phi v}\Big(\frac{\sqrt 3}{2\gamma} v\sin(2b)+\frac{\sqrt3 \kappa}{\alpha\sqrt\Delta}\phi\pi_\phi\Big)^2.
\end{equation}
As we will show in the next section, the dynamics driven by this effective Hamiltonian can be obtained analytically.

\section{The effective dynamics }\label{se:five}
To simplify the calculation of the dynamics determined by the effective Hamiltonian \eqref{Hamiltonian-e}, we choose a lapse function $N=v\phi/\alpha$, such that the effective Hamiltonian constraint can be reexpressed as
\begin{equation}
\begin{aligned}\label{eq:HamiltonianNH}
C=N\he=\phi^2v^2\sin^4(b)-\frac{1}{4\gamma^2}v^2\sin^2(2b)+\frac{1}{3+2\omega}\Big(\frac{\sqrt 3}{2\gamma} v\sin(2b)+\frac{\sqrt3 \kappa}{\alpha\sqrt\Delta}\phi\pi_\phi\Big)^2=0.
\end{aligned}
\end{equation}
Let $X=v\sin(2b)$, $Y=\phi\piphi$ and $Z=\phi v\sin^2(b)$. Then two constants of motion with respect to $C$ can be obtained as
\begin{equation}
\begin{aligned}
\xi_1&=\hbar X/4-Y,\\
\xi_2&=Z^2+AY^2+BXY,
\end{aligned}
\end{equation}
where
\begin{equation}\label{eq:ab}
\begin{aligned}
A&=\frac{8(3\omega+2)}{3\gamma^2(2\omega+3)\hbar^2},\\
B&=-\frac{4(\omega-1)}{\gamma^2(2\omega+3)\hbar}.
\end{aligned}
\end{equation}
Expressed by the two constants of motion, the constraint \eqref{eq:HamiltonianNH} can be reexpressed as
\begin{equation}\label{eq:xi}
\xi_2=\frac{8\omega}{\gamma^2(2\omega+3)\hbar^2}\xi_1^2.
\end{equation}
Thus, the Hamiltonian constraint \eqref{eq:HamiltonianNH} will be satisfied throughout the evolution as far as the two constants of motion are chosen such that Eq. \eqref{eq:xi} holds. The equations of motion for $X$, $Y$, and $Z$ can be easily derived by using Hamilton's equations with the Hamiltonian $C$, which, together with the Hamiltonian constraint \eqref{eq:HamiltonianNH}, leads to
\begin{equation}\label{eq:eom}
\begin{aligned}
\dot{Y}&=-2Z^2\\
Z^2&=-(A+\frac{4B}{\hbar})Y^2-\frac{4B}{\hbar}\xi_1Y+\xi_2=:\mathfrak{a}Y^2+\mathfrak{b}Y+\mathfrak{c},
\end{aligned}
\end{equation}
where we defined
\begin{equation}\label{eq:abc}
\begin{aligned}
\mathfrak{a}&=\frac{8(3\omega-8)}{3\gamma^2(2\omega+3)\hbar^2},\\
\mathfrak{b}^2-4\mathfrak{a}\mathfrak{c}&=\frac{256\xi_1^2}{3\gamma^4(2\omega+3)\hbar^4}.
\end{aligned}
\end{equation}
Thus the types of the solutions $Y(t)$ depend on the sign of $\mathfrak{b}^2-4\mathfrak{a}\mathfrak{c}$. For $\mathfrak{b}^2-4\mathfrak{a}\mathfrak{c}<0$, $Y(t)$ takes the form of a tangent function, while for $\mathfrak{b}^2-4\mathfrak{a}\mathfrak{c}>0$, it takes the form of a hyperbolic tangent function.
We are interested in the case with the coupling parameter $\omega\gg1$, which coincides with the Solar System experiments \cite{Will:2014kxa,will2018theory}. In this case Eq. \eqref{eq:abc} ensures that $\mathfrak{b}^2-4\mathfrak{a}\mathfrak{c}>0$. Then Eq. \eqref{eq:eom} gives
\begin{equation}\label{eq:doty2}
\dot{Y}=-2\mathfrak{a}(Y-y_1)(Y-y_2),
\end{equation}
where $y_1$ and $y_2$ ($y_1>y_2$) are the roots of the equation $\mathfrak{a}Y^2+\mathfrak{b}Y+\mathfrak{c}$=0. Thus the solutions to Eq. \eqref{eq:doty2} can be obtained as
\begin{equation}
Y_\pm(t)=y_1+\frac{y_2-y_1}{1\pm e^{2\mathfrak{a}(y_1-y_2)t}}.
\end{equation}
Taking account of the fact that $Z^2=\mathfrak{a}(Y-y_1)(Y-y_2)\geq 0$, we conclude the following two cases.
\begin{enumerate}[(i)]
\item For $\mathfrak{a}>0$, i.e., $\omega>8/3$, the solution is
\begin{equation}
Y_-(t)=\frac{3(\omega-1)-\sqrt{3(2\omega+3)}}{8-3\omega}\xi_1+\frac{2\sqrt{3(2\omega+3)}}{8-3\omega}\xi_1\left(1-e^{\frac{32\xi_1 t}{\gamma^2\hbar^2\sqrt{3(2\omega+3)}}}\right)^{-1}.
\end{equation}
\item For $\mathfrak{a}<0$, i.e., $-3/2<\omega<8/3$, the solution is
\begin{equation}
Y_+(t)=\frac{3(\omega-1)-\sqrt{3(2\omega+3)}}{8-3\omega}\xi_1+\frac{2\sqrt{3(2\omega+3)}}{8-3\omega}\xi_1\left(1+e^{\frac{32\xi_1 t}{\gamma^2\hbar^2\sqrt{3(2\omega+3)}}}\right)^{-1}.
\end{equation}
\end{enumerate}
By Eq. \eqref{eq:eom} we can obtain the expression of $Z_\pm(t)$ corresponding to $Y_\pm(t)$ as
\begin{equation}
\begin{aligned}
Z_-(t)&=\frac{2\sqrt{2}|\xi_1|}{\hbar\gamma\sqrt{3\omega-8}}\left|\sinh(\frac{16\xi_1}{\gamma^2\hbar^2\sqrt{3(2\omega+3)}}t)\right|^{-1},\\
Z_+(t)&=\frac{2\sqrt{2}|\xi_1|}{\hbar\gamma\sqrt{8-3\omega}}\left|\cosh(\frac{16\xi_1}{\gamma^2\hbar^2\sqrt{3(2\omega+3)}}t)\right|^{-1}.
\end{aligned}
\end{equation}
The equation of motion for $\phi$,  which can be derived by Hamilton's equation as well as the Hamiltonian constraint \eqref{eq:HamiltonianNH}, reads
\begin{equation}\label{eq:phi-solution}
\dot{\phi}_\pm=\frac{16\phi_\pm}{3\gamma^2(2\omega+3)\hbar^2}(5Y_\pm+3\xi_1).
\end{equation}
The solutions of Eq. \eqref{eq:phi-solution} can be obtained as
\begin{equation}
\begin{aligned}
\phi_-(t)=\phi_0\, 2^{\frac{5}{3\omega-8}}e^{-\frac{16\xi_1 t}{\gamma^2\hbar^2(3\omega-8)}}\left|\sinh(\frac{16\xi_1 t}{\sqrt{3(2\omega+3)}\gamma^2\hbar^2})\right|^{\frac{5}{3\omega-8}},\\
\phi_+(t)=\phi_0\, 2^{\frac{5}{3\omega-8}}e^{-\frac{16\xi_1 t}{\gamma^2\hbar^2(3\omega-8)}}\left|\cosh(\frac{16\xi_1 t}{\sqrt{3(2\omega+3)}\gamma^2\hbar^2})\right|^{\frac{5}{3\omega-8}},
\end{aligned}
\end{equation}
where $\phi_0$ is a integration constant. The dynamical evolution of $v$ and $b$ can be obtained by using the functions $X$, $Y$, $Z$, and $\phi$ as
\begin{equation}
v=\frac{\phi X^2}{4Z}+\frac{Z}{\phi},\ \sin(2b)=\frac{X}{v},\ \cos(2b)=1-\frac{2Z}{v\phi}.
\end{equation}
It should be noted that in the solutions obtained so far we adopted the coordinate time $t$ corresponding to the lapse function in Eq.~\eqref{eq:HamiltonianNH}. However, the Hubble parameter is defined with respect to the cosmological proper time $\tau$, which is related to the coordinate time by $\d\tau=8\pi G N\d t$. By denoting $\dot{v}:=\d v/\d t$, the Hubble parameter $\hh$ can be expressed as
\begin{equation}\label{Hubble}
  \hh=\frac{\alpha\dot{v}}{24\pi Gv^2\phi}=\frac{4\alpha\phi^2X\left(\dot{\phi} X Z+2Z\phi\dot{X}-\phi X\dot{Z}\right)+16\alpha Z^2\left(\dot{Z}\phi-Z\dot{\phi}\right)}{24\pi G\phi\left(\phi^2X^2+4Z^2\right)^2}.
\end{equation}
Taking account of the Solar System experiments, we consider the case $\omega>8/3$. In this case, the dynamics is described by the functions $Y_-(t),\ Z_-(t)$, and $\phi_-(t)$. Since the functions $Y_-(t)$ and $Z_-(t)$ are ill-defined at $t=0$, they are valid in the domain $t \in (-\infty,0)\cup(0,\infty)$, so is the lapse function $N=\phi v/\alpha$. Because $N$ does not vanish in this domain, as a time coordinate $t$ is well defined in each branch $(-\infty,0)$ or $(0,\infty)$. Moreover, for a given $t_0>0$, the integrals $\int_{\pm t_0}^{0^{\pm}}N(t)\d t$ and $\int_{\pm t_0}^{\pm\infty}N(t)\d t$ diverge. Hence the cosmological time $\tau$ ranges over $(-\infty,\infty)$ in either the branch of domain of $t$. Thus we can choose one of the branches, say $t\in (0,\infty)$, to cover the whole spacetime. Thanks to the divergence of the integrals, the hypersurfaces of $t=0$ and $t=\infty$ are actually the past and future timelike infinities respectively. Furthermore, the effective dynamics will return to the classical one for $v\gg 1$. This happens in the classical regions of $\frac{1}{t}\ll 1$ and $t\gg 1$ respectively.

\begin{figure}
\centering
\subfigure[]{
\begin{minipage}[t]{0.5\linewidth}
\centering
\includegraphics[width=\linewidth]{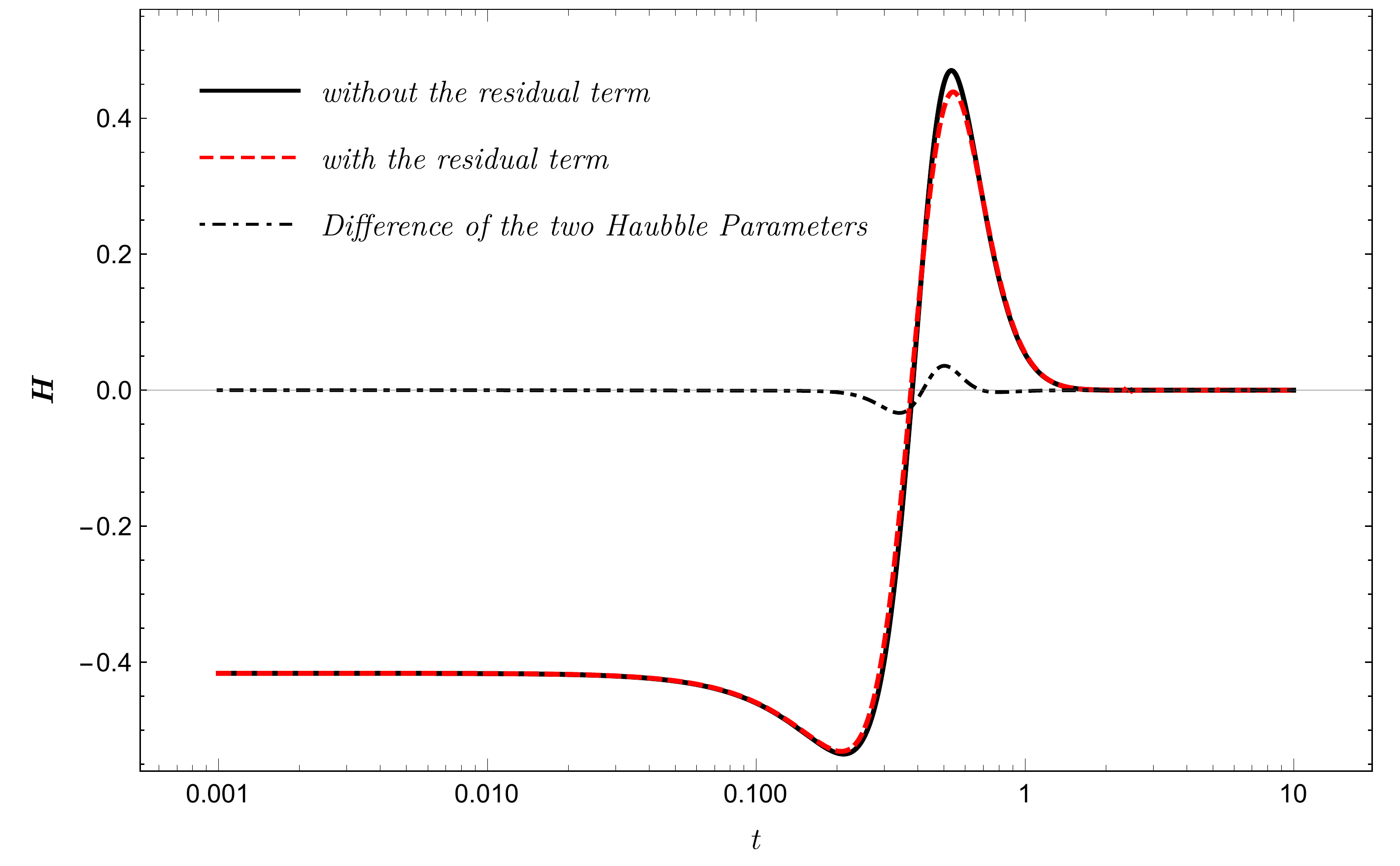}
\label{fig:Hubble}
\end{minipage}%
}%
\subfigure[]{
\begin{minipage}[t]{0.5\linewidth}
\centering
\includegraphics[width=\linewidth]{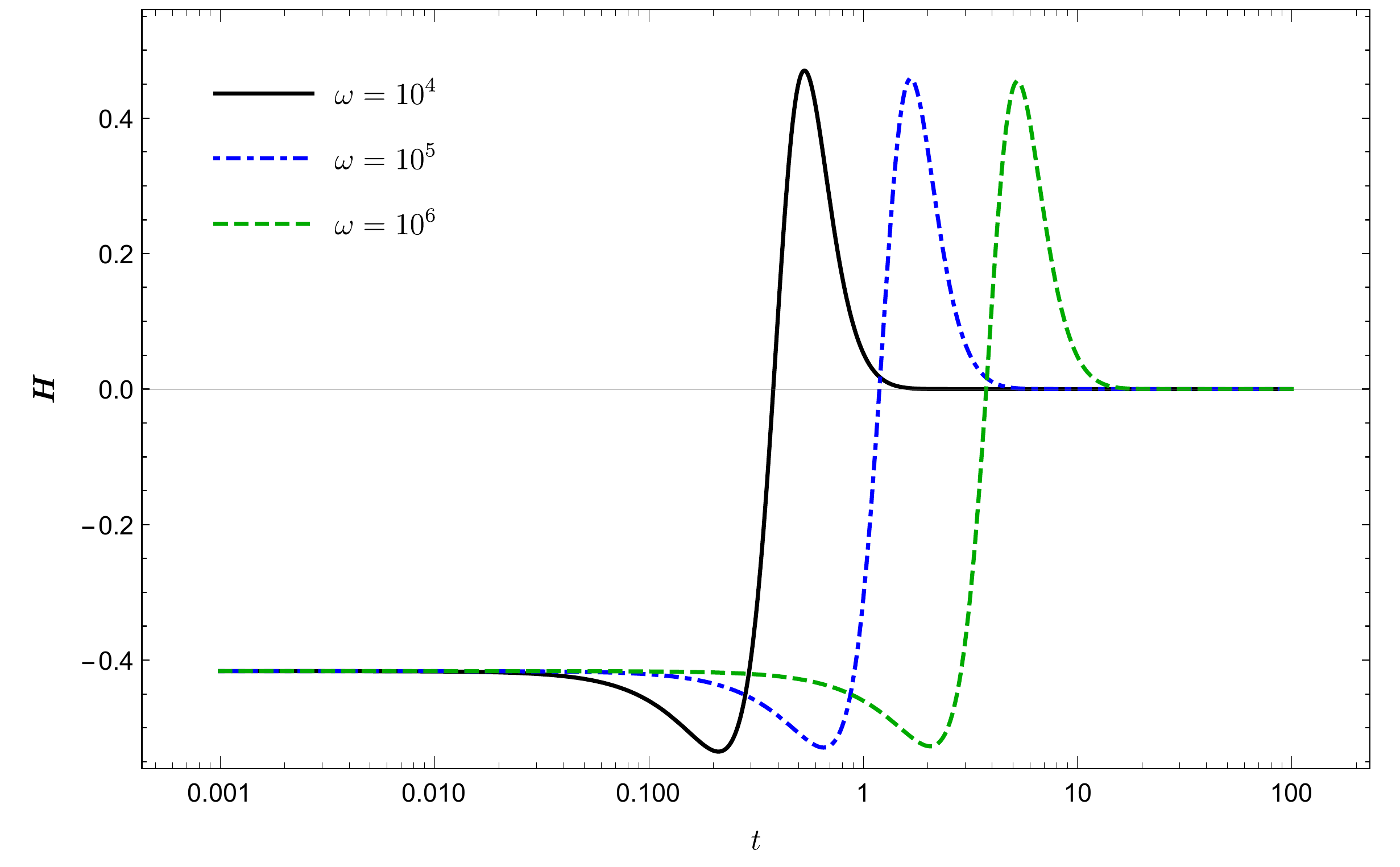}
\label{fig:omegaH}
\end{minipage}%
}
\centering
\caption{(a) Comparison of the evolutions of $\hh(t)$ driven by \eqref{Hamiltonian-e} (the solid line) and by \eqref{eq:effh} (the red dashed line): The difference between the two evolutions of $\hh(t)$ is also given (the black dot-dashed line). (b) Evolution of $\hh(t)$ with respect to different values of $\omega$.
The parameters in this plot are chosen as $\gamma =0.2357,\ \hbar =1$, $\ell_p =1$, $\xi =5$, and $\phi_0=1$ for both panels. In the left panel, we choose $\omega =10^4$. }\label{fig:hubble}
\end{figure}

\begin{figure}[h!tb]
\centering
\includegraphics[width=0.6\linewidth]{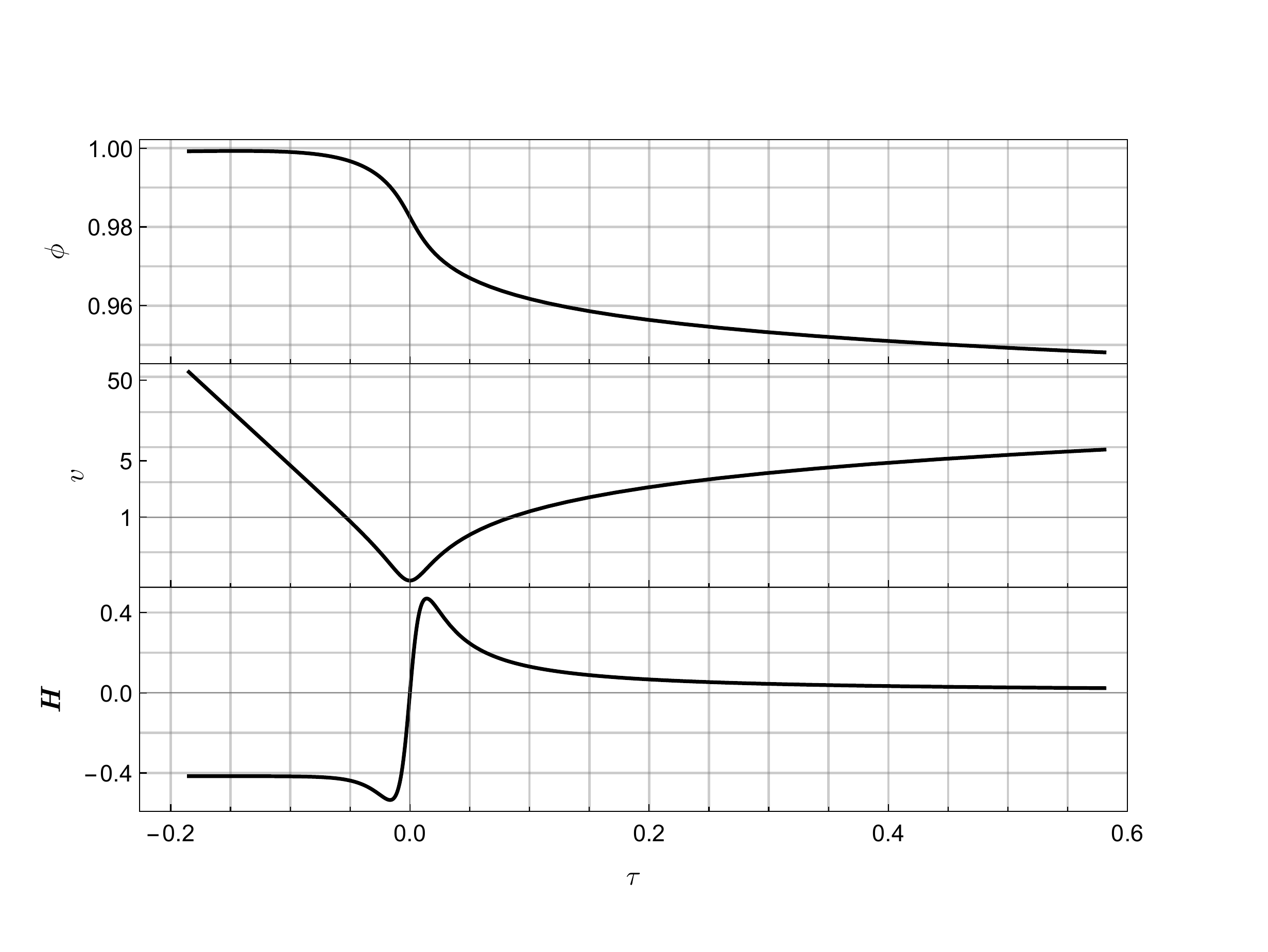}
\caption{The behaviors of $\phi$ and $v$ near the bounce compared with $\hh$. The parameters in the plot are chosen as $\gamma =0.2357,\ \hbar =1$, $\ell_p =1$, $\xi =5$, $\phi_0=1$, and $\omega =10^4$.}
\label{fig:vH}
\end{figure}

Now we consider the dynamical behavior of the Universe with $t\in (0,\infty)$. As $t\rightarrow0$, the leading terms of the functions $Y_-,Z_-,X_-,$ and $\phi_-$ read respectively
\begin{equation}\label{eq:fiveone}
\begin{aligned}
  Y_{-}(t)&\cong \frac{3\gamma^2\hbar^2(2\omega+3)}{16(3\omega-8)}\,\frac{1}{t},\\
  Z_{-}(t)&\cong\frac{\gamma\hbar\sqrt{6(2\omega+3)}}{8\sqrt{3\omega-8}}\,\frac{1}{|t|},\\
  X_{-}(t)&\cong-\frac{3\gamma^2\hbar(2\omega+3)}{4(3\omega-8)}\,\frac{1}{t},\\
  \phi_{-}(t)&\cong\phi_0\left(\frac{32\xi_1}{\sqrt{3(2\omega+3)}\gamma^2\hbar^2}\right)^{\frac{5}{3\omega-8}} t^{5/(3\omega-8)}.
\end{aligned}
\end{equation}
Thus, their derivatives with respect to $t$ are respectively
\begin{equation}
  \begin{aligned}
   \dot{Y}_{-}(t)&\cong-\frac{1}{t}Y_{-}(t),\quad  &\dot{Z}_{-}(t)\cong-\frac{1}{t}Z_{-}(t),\\
   \dot{X}_{-}(t)&\cong-\frac{1}{t}X_{-}(t),\quad &\dot{\phi}_{-}(t)\cong\frac{5}{3\omega-8}\,\frac{1}{t}\phi_{-}(t).
  \end{aligned}
\end{equation}
Hence, as $t\rightarrow0$, by Eq. \eqref{Hubble} the Hubble parameter approaches
\begin{align}\label{eq:Hubble0}
  \hh&\cong-\frac{8\alpha(\omega-1)}{8\pi\gamma\ell_p^2\sqrt{6(3\omega-8)(2\omega+3)}}<0.
\end{align}
Let us consider the other side. As $t\to\infty$, the leading terms of those functions become respectively
\begin{equation}
\begin{aligned}
  Y_{-}(t)&\cong\frac{3(\omega-1)- \sgn(t\xi_1)\sqrt{3(2\omega+3)}}{8-3\omega}\xi_{1},\\
  X_{-}(t)&\cong\frac{4\left(5- \sgn(t\xi_1)\sqrt{3(2\omega+3)}\right)}{(8-3\omega)\hbar}\xi_1,\\
  Z_{-}(t)&\cong\frac{2\sqrt{2}|\xi_{1}|}{\hbar\gamma\sqrt{3\omega-8}} \exp\left(\frac{-16\left|\xi_{1}t\right|}{\gamma^{2}\hbar^{2}\sqrt{3(2\omega+3)}}\right),\\
  \phi_{-}(t)&\cong\exp\left[\frac{16\xi_1 t}{\gamma^2\hbar^2(3\omega-8)}\left(\frac{5\,\sgn(t\xi_1)}{\sqrt{3(2\omega+3)}}-1\right)\right].
\end{aligned}
\end{equation}
Then their time derivatives are respectively
\begin{equation}
  \begin{aligned}
    \dot{Y}_{-}(t)&\cong 0,\\
    \dot{X}_{-}(t)&\cong 0,\\
    \dot{Z}_{-}(t)&\cong- \sgn(t\xi_1)\frac{16\xi_1}{\gamma^2\hbar^2\sqrt{3(2\omega+3)}}Z_-(t),\\
    \dot{\phi}_{-}(t)&\cong\frac{16\xi_1 }{\gamma^2\hbar^2(3\omega-8)}\left(\frac{5\,\sgn(t\xi_1)}{\sqrt{3(2\omega+3)}}-1\right)\phi_{-}(t).
  \end{aligned}
\end{equation}
Hence the asymptotic behavior of the Hubble parameter for $t\to \infty$ reads
\begin{align}\label{eq:Hubblei}
  \hh\cong\lim_{t\to\infty} \frac{256 \alpha\xi^2    e^{-\frac{16 |\xi_1  t|}{\gamma ^2 \sqrt{6 \omega +9} \hbar ^2}}}{24\pi G \gamma ^3 \hbar ^3 \sqrt{3 \omega -8} \sqrt{3 \omega +\frac{9}{2}}} = 0.
\end{align}
Equations \eqref{eq:Hubble0} and \eqref{eq:Hubblei} imply that there exists at lease one moment $t_0\in(0,\infty)$ such that $H(t_0)=0$. Hence a bounce of the Universe may happen at $t=t_0$. On one side, the negative Hubble constant around $t=0^+$ implies that the Universe goes through an asymptotical de Sitter epoch there. On the other side, the fact that $\hh(t)$ approaches to $0^+$ as $t\to\infty$ implies that the effective theory returns to the classical Brans-Dicke cosmology at late time. It is easy to check that the asymptotic behavior of the Universe would not change if the residual term in the effective Hamiltonian \eqref{eq:effh} was taken into account. However, the detailed evolution around the bounce would be influenced by that term.
The numerical simulation for the evolution of the Hubble parameter is plotted in Fig. \ref{fig:hubble}. In the left panel, the dynamics of $\hh(t)$ driven by the Hamiltonian constraints \eqref{eq:effh} and \eqref{Hamiltonian-e} are compared. In the right panel, the dynamics of $\hh(t)$ driven by the Hamiltonian constraint  \eqref{Hamiltonian-e} with respect to different values of $\omega$ are shown. According to the results, there is only a single bounce with $\hh(t)=0$. Around the bounce, the residual term does affect the dynamics. However,
for various values of $\omega$, the qualitative features of $\hh(t)$ are not influenced. Furthermore, the evolutions of $\phi$ and $v$ with respect to the cosmological time $\tau$ are also plotted in Fig.~\ref{fig:vH}. As shown in this plot, $v(\tau)$ bounces at $\tau_0$ with $\hh(\tau_0)=0$. In the de Sitter epoch, $v(\tau)$ grows exponentially as $\tau$ goes from $0$ to $-\infty$. It is straightforward to check that the dynamics of $\hh(t)$, $\phi(t)$, and $v(t)$ for $t\in (-\infty,0)$ behaves similar to that for $t\in (0,\infty)$.

\section{Discussion}\label{se:six}

In the previous sections, to inherit more features of full LQBDT, we dealt with the Euclidean and Lorentzian terms of the Hamiltonian constraint independently in LQBDC. The Hamiltonian constraint operator \eqref{eq:cc} alternative to the one obtained in \cite{zhang2013loop} was constructed in Sec. \ref{se:three}.
The effective Hamiltonian constraint \eqref{eq:effh} was also derived from the alternative Hamiltonian operator by the semiclassical analysis in Sec. \ref{se:four}. It turns out that there exists a residual quantum correction term in the effective Hamiltonian, which could not be obtained simply by replacing $b\to \sin(b)$ or $b\to \sin(2b)/2$ in the classical Hamiltonian constraint. This is a particular property of our LQBDC.
The dynamics given by the effective Hamiltonian constraint was analyzed in Sec. \ref{se:five}. The evolution equation of the Universe was solved analytically by getting rid of the residual term which is of $\hbar^2$-order. The dynamical behaviors of the Hubble parameter for the physically interesting case of $\omega\gg 1$ was considered. It turns out that the classical singularity is resolved by a quantum bounce which relates a de Sitter epoch to a usual classical Brans-Dicke cosmology. Both the evolutions driven by the effective Hamiltonian \eqref{Hamiltonian-e} and by the original \eqref{eq:effh} with the residual term were numerically computed and plotted in Fig.~\ref{fig:Hubble}. The comparison of the two evolutions shows that the two Hamiltonians determine the qualitatively same dynamics. However, the residual term affected the evolution around the bounce, while they give the same asymptotic behaviors.

Since an asymptotical de Sitter epoch appears in our cosmological model, it is interesting to see whether that epoch of the model can match the observation of current accelerating Universe. By substituting \eqref{eq:fiveone} into \eqref{Hubble}, the Hubble parameter in the asymptotical de Sitter epoch can be expressed as
\begin{equation}
\hh(t)\cong -\frac{2\alpha}{\sqrt{6}\pi \gamma\ell_p^2} \sqrt{\frac{3 \omega -8}{2 \omega +3}}\, \frac{2 (\omega -1) (3 \omega -8)+ (2 \omega +3) (3 \omega -13)\gamma ^2\phi_-(t)^2  }{ \left(\,2(3 \omega -8)+3(2 \omega +3)\gamma ^2 \phi_-(t)^2\, \right)^2}.
\end{equation}
Hence, if one asked $\hh(t)$ at some fixed $t$ to match the observation, the value of $\phi_-(t)$ would have to be sufficiently large. For instance, letting $\omega=10^4$, one has $\phi_-(t)=8.899\times 10^{30}$. Moreover, $\hh(t)$ should change slowly at the moment $t$. Such a requirement could be achieved by choosing $\phi_0$ and $\xi_1$ in the expression of $\phi_-(t)$ properly. However, it is straightforward to check that in this case, the effective gravitational constant $G/\phi_-(t)$ in the Brans-Dicke theory is far away from the observational value because of the huge value of $\phi_-(t)$. Thus there is no evidence that the emerged asymptotical de Sitter epoch could match our current Universe.

\section*{Acknowledgements}
The authors would like to thank Chun-Yen Lin for discussion. This work is supported by NSFC with Grants No. 11875006 and No. 11961131013. C. Z. acknowledges the support by the Polish Narodowe Centrum Nauki, Grant No. 2018/30/Q/ST2/00811.


\begin{thebibliography}{34}
\providecommand{\natexlab}[1]{#1}
\providecommand{\url}[1]{\texttt{#1}}
\expandafter\ifx\csname urlstyle\endcsname\relax
  \providecommand{\doi}[1]{doi: #1}\else
  \providecommand{\doi}{doi: \begingroup \urlstyle{rm}\Url}\fi

\bibitem[Ashtekar and Lewandowski(2004)]{ashtekar2004back}
A.~Ashtekar and J.~Lewandowski.
\newblock Background independent quantum gravity: a status report.
\newblock \emph{Classical and Quantum Gravity}, 21\penalty0 (15):\penalty0 R53,
  2004.

\bibitem[Rovelli(2005)]{rovelli2005quantum}
C.~Rovelli.
\newblock \emph{quantum gravity}.
\newblock Cambridge University Press, 2005.

\bibitem[Thiemann(2007)]{thiemann2007modern}
T.~Thiemann.
\newblock \emph{Modern canonical quantum general relativity}.
\newblock Cambridge University Press, 2007.

\bibitem[Han et~al.(2007)Han, Ma, and Huang]{han2007fundamental}
M.~Han, Y.~Ma, and W.~Huang.
\newblock Fundamental structure of loop quantum gravity.
\newblock \emph{International Journal of Modern Physics D}, 16\penalty0
  (09):\penalty0 1397--1474, 2007.

\bibitem[Rovelli and Smolin(1994)]{rovelli1994physical}
C.~Rovelli and L.~Smolin.
\newblock The physical hamiltonian in nonperturbative quantum gravity.
\newblock \emph{Physical review letters}, 72\penalty0 (4):\penalty0 446, 1994.

\bibitem[Ashtekar and Lewandowski(1997{\natexlab{a}})]{ashtekar1997quantum}
A.~Ashtekar and J.~Lewandowski.
\newblock Quantum theory of geometry: I. area operators.
\newblock \emph{Classical and Quantum Gravity}, 14\penalty0 (1A):\penalty0 A55,
  1997{\natexlab{a}}.

\bibitem[Ashtekar and Lewandowski(1997{\natexlab{b}})]{ashtekar1997quantumII}
A.~Ashtekar and J.~Lewandowski.
\newblock Quantum theory of geometry ii: Volume operators.
\newblock \emph{Advances in Theoretical and Mathematical Physics}, 1\penalty0
  (2):\penalty0 388--429, 1997{\natexlab{b}}.

\bibitem[Thiemann(1998{\natexlab{a}})]{thiemann1998length}
T.~Thiemann.
\newblock A length operator for canonical quantum gravity.
\newblock \emph{Journal of Mathematical Physics}, 39\penalty0 (6):\penalty0
  3372--3392, 1998{\natexlab{a}}.

\bibitem[Ma et~al.(2010)Ma, Soo, and Yang]{ma2010new}
Y.~Ma, C.~Soo, and J.~Yang.
\newblock New length operator for loop quantum gravity.
\newblock \emph{Physical Review D}, 81\penalty0 (12):\penalty0 124026, 2010.

\bibitem[Yang and Ma(2016)]{yang2016new}
J.~Yang and Y.~Ma.
\newblock New volume and inverse volume operators for loop quantum gravity.
\newblock \emph{Phys. Rev. D}, 94:\penalty0 044003, Aug 2016.
\newblock \doi{10.1103/PhysRevD.94.044003}.
\newblock URL \url{https://link.aps.org/doi/10.1103/PhysRevD.94.044003}.

\bibitem[Thiemann(1998{\natexlab{b}})]{thiemann1998quantum}
T.~Thiemann.
\newblock Quantum spin dynamics (qsd).
\newblock \emph{Classical and Quantum Gravity}, 15\penalty0 (4):\penalty0 839,
  1998{\natexlab{b}}.

\bibitem[Thiemann(2006)]{thiemann2006phoenix}
T.~Thiemann.
\newblock The phoenix project: master constraint programme for loop quantum
  gravity.
\newblock \emph{Classical and Quantum Gravity}, 23\penalty0 (7):\penalty0 2211,
  2006.

\bibitem[Han and Ma(2006)]{han2006master}
M.~Han and Y.~Ma.
\newblock Master constraint operators in loop quantum gravity.
\newblock \emph{Physics Letters B}, 635\penalty0 (4):\penalty0 225--231, 2006.

\bibitem[Yang and Ma(2015)]{yang2015new}
J.~Yang and Y.~Ma.
\newblock New hamiltonian constraint operator for loop quantum gravity.
\newblock \emph{Physics Letters B}, 751:\penalty0 343--347, 2015.

\bibitem[Domagala(2015)]{Domagala2015kse}
M.~Domagala.
\newblock \emph{{On quantum model of the masless Klein-Gordon field coupled to
  gravity}}.
\newblock PhD thesis, Warsaw U., 2015.
\newblock URL \url{https://depotuw.ceon.pl/handle/item/1147}.

\bibitem[Alesci et~al.(2015)Alesci, Assanioussi, Lewandowski, and
  M\"akinen]{alesci2015hamiltonian}
E.~Alesci, M.~Assanioussi, J.~Lewandowski, and I.~M\"akinen.
\newblock Hamiltonian operator for loop quantum gravity coupled to a scalar
  field.
\newblock \emph{Phys. Rev. D}, 91:\penalty0 124067, Jun 2015.
\newblock \doi{10.1103/PhysRevD.91.124067}.

\bibitem[Bonzom and Freidel(2011)]{bonzom2011hamiltonian}
V.~Bonzom and L.~Freidel.
\newblock The hamiltonian constraint in 3d riemannian loop quantum gravity.
\newblock \emph{Classical and Quantum Gravity}, 28\penalty0 (19):\penalty0
  195006, 2011.

\bibitem[Alesci et~al.(2013)Alesci, Liegener, and Zipfel]{alesci2013matrix}
E.~Alesci, K.~Liegener, and A.~Zipfel.
\newblock Matrix elements of lorentzian hamiltonian constraint in loop quantum
  gravity.
\newblock \emph{Physical Review D}, 88\penalty0 (8):\penalty0 084043, 2013.

\bibitem[Zhang et~al.(2018)Zhang, Lewandowski, and Ma]{zhang2018towards}
C.~Zhang, J.~Lewandowski, and Y.~Ma.
\newblock Towards the self-adjointness of a hamiltonian operator in loop
  quantum gravity.
\newblock \emph{Physical Review D}, 98\penalty0 (8):\penalty0 086014, 2018.

\bibitem[Zhang et~al.(2019)Zhang, Lewandowski, Li, and Ma]{zhang2019bouncing}
C.~Zhang, J.~Lewandowski, H.~Li, and Y.~Ma.
\newblock Bouncing evolution in a model of loop quantum gravity.
\newblock \emph{Physical Review D}, 99\penalty0 (12):\penalty0 124012, 2019.

\bibitem[Bojowald(2001)]{bojowald2001absence}
M.~Bojowald.
\newblock Absence of a singularity in loop quantum cosmology.
\newblock \emph{Physical Review Letters}, 86\penalty0 (23):\penalty0 5227,
  2001.

\bibitem[Ashtekar et~al.(2006)Ashtekar, Pawlowski, and
  Singh]{ashtekar2006quantumnature}
A.~Ashtekar, T.~Pawlowski, and P.~Singh.
\newblock Quantum nature of the big bang: Improved dynamics.
\newblock \emph{Phys. Rev. D}, 74:\penalty0 084003, Oct 2006.
\newblock \doi{10.1103/PhysRevD.74.084003}.

\bibitem[Ding et~al.(2009)Ding, Ma, and Yang]{ding2009effective}
Y.~Ding, Y.~Ma, and J.~Yang.
\newblock Effective scenario of loop quantum cosmology.
\newblock \emph{Physical review letters}, 102\penalty0 (5):\penalty0 051301,
  2009.

\bibitem[Brans and Dicke(1961)]{brans1961mach}
C.~Brans and R.~H. Dicke.
\newblock Mach's principle and a relativistic theory of gravitation.
\newblock \emph{Physical review}, 124\penalty0 (3):\penalty0 925, 1961.

\bibitem[Zhang and Ma(2011)]{Zhang:2011vg}
X.~Zhang and Y.~Ma.
\newblock {Nonperturbative Loop Quantization of Scalar-Tensor Theories of
  Gravity}.
\newblock \emph{Phys. Rev. D}, 84:\penalty0 104045, 2011.
\newblock \doi{10.1103/PhysRevD.84.104045}.

\bibitem[Zhang et~al.(2013)Zhang, Artymowski, and Ma]{zhang2013loop}
X.~Zhang, M.~Artymowski, and Y.~Ma.
\newblock Loop quantum brans-dicke cosmology.
\newblock \emph{Physical Review D}, 87\penalty0 (8):\penalty0 084024, 2013.

\bibitem[Artymowski et~al.(2013)Artymowski, Ma, and Zhang]{Artymowski:2013qua}
M.~Artymowski, Y.~Ma, and X.~Zhang.
\newblock {Comparison between Jordan and Einstein frames of Brans-Dicke gravity
  a la loop quantum cosmology}.
\newblock \emph{Phys. Rev. D}, 88\penalty0 (10):\penalty0 104010, 2013.
\newblock \doi{10.1103/PhysRevD.88.104010}.

\bibitem[Yang et~al.(2009)Yang, Ding, and Ma]{yang2009alternative}
J.~Yang, Y.~Ding, and Y.~Ma.
\newblock Alternative quantization of the hamiltonian in loop quantum
  cosmology.
\newblock \emph{Physics Letters B}, 682\penalty0 (1):\penalty0 1--7, 2009.

\bibitem[Assanioussi et~al.(2018)Assanioussi, Dapor, Liegener, and
  Paw{\l}owski]{assanioussi2018emergent}
M.~Assanioussi, A.~Dapor, K.~Liegener, and T.~Paw{\l}owski.
\newblock Emergent de sitter epoch of the quantum cosmos from loop quantum
  cosmology.
\newblock \emph{Physical review letters}, 121\penalty0 (8):\penalty0 081303,
  2018.

\bibitem[Li et~al.(2018)Li, Singh, and Wang]{li2018towards}
B.-F. Li, P.~Singh, and A.~Wang.
\newblock Towards cosmological dynamics from loop quantum gravity.
\newblock \emph{Physical Review D}, 97\penalty0 (8):\penalty0 084029, 2018.

\bibitem[Ashtekar et~al.(2003)Ashtekar, Bojowald, Lewandowski,
  et~al.]{ashtekar2003mathematical}
A.~Ashtekar, M.~Bojowald, J.~Lewandowski, et~al.
\newblock Mathematical structure of loop quantum cosmology.
\newblock \emph{Advances in Theoretical and Mathematical Physics}, 7\penalty0
  (2):\penalty0 233--268, 2003.

\bibitem[Assanioussi et~al.(2017)Assanioussi, Lewandowski, and
  M{\"a}kinen]{assanioussi2017time}
M.~Assanioussi, J.~Lewandowski, and I.~M{\"a}kinen.
\newblock Time evolution in deparametrized models of loop quantum gravity.
\newblock \emph{Physical Review D}, 96\penalty0 (2):\penalty0 024043, 2017.

\bibitem[Will(2014)]{Will:2014kxa}
C.~M. Will.
\newblock {The Confrontation between General Relativity and Experiment}.
\newblock \emph{Living Rev. Rel.}, 17:\penalty0 4, 2014.
\newblock \doi{10.12942/lrr-2014-4}.

\bibitem[Will(2018)]{will2018theory}
C.~M. Will.
\newblock \emph{Theory and experiment in gravitational physics}.
\newblock Cambridge university press, 2018.

\end{thebibliography}

\end{document}